\documentclass[aps,prd,reprint,superscriptaddress]{revtex4-1}

\usepackage{amsmath}
\usepackage{amsfonts}
\usepackage{graphicx}
\usepackage{mathrsfs}
\usepackage{comment}
\usepackage{verbatim}   
\usepackage{color}      
\usepackage{subfigure}  
\usepackage{hyperref}   
\newcommand{\Slash}[1]{{\ooalign{\hfil#1\hfil\crcr\raise.167ex\hbox{/}}}}

\newcommand{\beq}{\begin{equation}}  \newcommand{\eeq}{\end{equation}}
\newcommand{\bef}{\begin{figure}}  \newcommand{\eef}{\end{figure}}
\newcommand{\bec}{\begin{center}}  \newcommand{\eec}{\end{center}}
\newcommand{\non}{\nonumber}  
\newcommand{\laq}[1]{\label{eq:#1}}  

\newcommand{\Eq}[1]{Eq.~(\ref{eq:#1})}

\newcommand{\eq}[1]{(\ref{eq:#1})}

\newcommand{\ab}[1]{\left|{#1}\right|}
\newcommand{\vev}[1]{ \left\langle {#1} \right\rangle }

\newcommand{\lac}[1]{\label{chap:#1}}
\newcommand{\U}[1]{{\rm U{#1} } }
\newcommand{\SU}[1]{{\rm SU{#1} } }

\def\o{\over}
\def\a{\alpha}

\def\d{\delta}
\def\e{\epsilon}

\def\p{\psi}

\def\t{\tau}

\def\L{\Lambda}
\def\F{\Phi}

\def\tl{\tilde}
\def\*{\dagger}

\def\({\left(}
\def\){\right)}

\def\O{\mathcal{O}}

\def\tr{\mathop{\rm tr}}

\newcommand{\OR}{~{\rm or}~}
\newcommand{\AND}{~{\rm and}~}

\newcommand{\GEV}{ {\rm ~GeV} }
\newcommand{\TEV}{ {\rm ~TeV} }

\begin{document}

\title{\boldmath Peccei-Quinn symmetry from a hidden gauge group structure}

\author{Hye-Sung Lee} 
\author{Wen Yin} 
\affiliation{Department of Physics, Korea Advanced Institute of Science and Technology, Daejeon 34141, Korea}

\begin{abstract}
We introduce a natural origin of the Peccei-Quinn (PQ) symmetry with a sufficiently good precision.
In the standard model, the baryon number symmetry $\U(1)_B$ arises accidentally due to the $\SU(3)_C$ color gauge symmetry, and it protects the proton from a decay at a sufficient level.
Likewise, if there is an $\SU(N)$ gauge symmetry in the hidden sector, an accidental hidden baryon number symmetry $\U(1)_{B_H}$ can appear.
The hidden baryon number is solely obtained by the structure of the $\SU(N)$ group.
In particular, the quality of the $\U(1)_{B_H}$ can be arbitrarily good for an asymptotically-free theory with large enough $N$.  The $\U(1)_{B_H}$ can be identified as a PQ symmetry.
Using our findings, we build two types of novel composite axion models: a model where only one $\SU(N)$ gauge symmetry is required to both guarantee the quality and break the $\U(1)_{B_H}$, and a model with $\SU(N)\times\SU(M)$ gauge symmetry where the exotic quarks responsible to the axion-gluon coupling do not confine into exotic hadrons through the dynamical breaking of the PQ symmetry, and have masses of TeV scales.
\end{abstract}

\maketitle

\section{Introduction}
Peccei-Quinn (PQ) symmetry, a global $\U(1)_{PQ}$, is a leading candidate to solve the strong CP problem of the standard model (SM) \cite{Peccei:1977hh,Peccei:1977ur}, and axion is a pseudo Nambu-Goldstone boson (pNGB) of the $\U(1)_{PQ}$ \cite{Weinberg:1977ma,Wilczek:1977pj,Kim:1979if,Shifman:1979if,Dine:1981rt,Zhitnitsky:1980tq}.
The $\U(1)_{PQ}$ carries the quantum anomaly to the color gauge symmetry $\SU(3)_C$ of the SM, and thus the axion gets a potential with a CP-conserving minimum due to the non-perturbative effect of the QCD. 
As a pleasant surprise, the axion can also explain the dark matter puzzle~\cite{Abbott:1982af,Dine:1982ah,Preskill:1982cy}.
(See e.g. Refs.~\cite{Kim:2008hd,Wantz:2009it,Ringwald:2012hr,Kawasaki:2013ae,Marsh:2015xka} for recent reviews.)

The implicit assumption is that the $\U(1)_{PQ}$ should have a very good quality at the perturbative level.
However, it is believed that any global symmetry should be explicitly broken by Planck-scale physics.
(See e.g. Refs.~\cite{Misner:1957mt,Banks:2010zn}.)
Thus, the PQ symmetry should not be imposed by hand but be an accidental symmetry of the Lagrangian.  
Within the $4D$ quantum field theory, it was pointed out that a PQ symmetry with good enough quality can be obtained by imposing discrete gauge symmetries~\cite{Chun:1992bn,BasteroGil:1997vn,Babu:2002ic,Dias:2002hz, Harigaya:2013vja}, abelian gauge symmetries~\cite{Fukuda:2017ylt,Duerr:2017amf,Bonnefoy:2018ibr}, and non-abelian gauge symmetries~\cite{Randall:1992ut,DiLuzio:2017tjx, Lillard:2018fdt}.

Within the SM, we can actually find an accidental $\U(1)$ symmetry, the baryon number symmetry $\U(1)_B$, which is of a sufficient precision to suppress the proton decay.
The existence of the $\U(1)_B$ can be understood from the triality (or $Z_3$ symmetry) of the $\SU(3)_C$ gauge group~\cite{Georgi:1999wka}, where all quarks are charged.
From the triality, the baryon number is conserved among all possible renormalizable terms if there is no scalar charged under the $\SU(3)_C.$ 
In the SM, the quarks are chiral under the SM gauge group ${\cal G}_{\rm SM}\equiv \SU(3)_C\times \SU(2)_L \times \U(1)_Y,$ and the $\U(1)_B$ is anomalous to the $\SU(2)_L\times \U(1)_Y$.
This leads us to contemplate that the $\U(1)_{PQ}$ may have a similar origin as the $\U(1)_B$.

In this article, we show that if there is an $\SU(N)$ gauge theory, a hidden baryon number can be generally and solely obtained from the structure of the $\SU(N)$ gauge group, 
and can be a good candidate of the $\U(1)_{PQ}$ charge. 
The approximate conservation of the hidden baryon number 
is due to the ``$N$-ality'' of the $\SU(N)$ gauge group just like {the $\U(1)_B$ case.}
In particular, if one considers an asymptotically-free $\SU(N)$ gauge theory with large enough $N$, the hidden baryon number symmetry $\U(1)_{B_H}$ is arbitrarily precise. 
The $\U(1)_{B_H}$ will be anomalous to the $\SU(3)_C$ with chiral fermions under $\SU(N)\times \SU(3)_C$, and thus can be identified as the $\U(1)_{PQ}$ that solves the strong CP problem.

We also build concrete models to demonstrate our idea. 
Interestingly, due to the chiral fermions under $\SU(N)\times {\cal G}_{\rm SM}$, light exotic quarks and leptons can exist that can be tested in collider experiments.

\section{\boldmath Hidden baryon number in $\SU(N)$ gauge theories}
Here, we study the possibility of the $\U(1)_{PQ}$ as a hidden baryon number symmetry $\U(1)_{B_H}$ from an $\SU(N)$ gauge theory.
{We first review the baryon number in the QCD. Then we discuss that for an $\SU(N)$ gauge theory, in general there is a hidden baryon number solely determined by the group structure, {which can be conserved} at a good precision thanks to the ``$N$-ality".}

We consider the $\SU(N)$ QCD with flavor $F>1$, which has $F$ Weyl fermions, $q^{\a_i}_i$ and $\overline{q^{\a_i}}_i$ of fundamental and anti-fundamental representations, where $i$ ($\a_i$) is the flavor (gauge) indices. 
If $N=3$, the model corresponds to the QCD of the SM. The Lagrangian of a general form for $N={\rm even}$ (for the sake of simplicity of discussion) is 
\beq
{\cal L} = {\cal L}_{ n=0} + M_{\rm pl}^{-3N/2+4} \big(g_1 q^{N}+g_2 \bar{q}^{{N}}\big) + \cdots + h.c.
\eeq
where $g_{1}$, $g_2$ are dimensionless parameters with flavor indices omitted; ${\cal L}_{n=0}= M_{ij} \bar{q}_i q_j + \cdots$ has operators with equal numbers of $q$ and $\bar{q}$; we neglect higher dimensional terms in ``$\cdots$'' for an illustrative purpose; $q^{N}\equiv\epsilon_{\a_1 \cdots \a_N} q_{i_1}^{\a_1} \cdots q_{i_N}^{\a_N}$ is a gauge singlet with antisymmetric contraction of indices. 

We notice that in each term the power of the quarks always differs from that of the antiquarks by $n \times N$ where $n$ is an integer.  
The operators in ${\cal L}_{n=0}$ have a baryon number symmetry $\U(1)_{B}$ with the charge
\beq
q_i : 1, \qquad \bar{q}_i : -1
\eeq
up to a normalization factor.
However, the operator with $n \neq 0$ has at least $n N$ (anti)quarks (e.g. the second terms), and thus breaks the $\U(1)_{B}$.
The dimension of the operator satisfies
\beq
\laq{fd}
d \geq {3N\over 2} .
\eeq
When $N$ is odd, $q^N$ is not a Lorentz invariant and $d$ should be larger. 
For the $F=1$ case, the breaking term has larger dimension because the leading breaking operator has $q^N$ and derivatives.

In a general $\SU(N)$ gauge theory, there can be multiple representations of fermions and bosons. 
($N$ can be either even or odd.)
Even in this case, $\U(1)_{B_H}$ can exist and be precise due to the ``$N$-ality" of the group theory~\cite{Georgi:1999wka}, which is an extension of the triality of the $\SU(3)_C$. 

A scalar or fermion, $\F_i$ of an $\SU(N)$ can have a general representation, $r_i$ decomposed by specific positive integers, $(n_i, \bar{m}_i)$ of fundamental and anti-fundamental representations. The $n_i$ and $\bar{m}_i$ can be understood as the numbers of fundamental and anti-fundamental indices of a representation, respectively. We take a convention that $|n_i +\bar{m}_i|$ is minimized by the contractions of indices with $\e_{\a_1 \cdots \a_N}$. (For $n_i, \bar{m}_i \gtrsim N$, this convention may not be unique, but it does not change our discussion.)  
For instance, a fundamental representation, $r_i=N$ has $(1, 0)$, a second rank symmetric (or antisymmetric) representation, $r_i=N(N+1)/2$ [or $N(N-1)/2)$] has $(2, 0)$, and an adjoint representation or singlet, $r_i=N^2-1$ has $(1, 1)$, etc.
A complex-conjugate representation, $\bar{r}_i$ has $(\bar{m}_i, n_i)$.

$n_i-\bar{m}_i~ ({\rm mod}~{N}) $ is noted as ``$N$-ality", which is conserved in the contraction of the gauge indices. 
In particular, a gauge singlet operator (with various indices omitted),
\beq
\laq{op}
{O}_n= {\prod_{i=1}^\ell {\F_i }},
\eeq 
which is a product of fields of general representations has vanishing ``$N$-ality'':
\beq
\laq{vnN}
\sum_{i=1}^{\ell}  n_i - \bar{m}_i = n N .
\eeq
Here, the subscript $n$ of $O_n$ represents the $n$ in Eq.~\eq{vnN}.
$\ell$ is the total number of fields in the operator, and $\F_i$ and $\F_j$ could be the same field for $i\neq j$.

One finds that  $O_n$ with $n=0$ conserves a $\U(1)$ symmetry, which is the $\U(1)_{B_H}$ with the hidden baryon number $b_H(r_i)$ as 
\beq 
\laq{rule}
\F_i : n_i-\bar{m}_i 
\eeq
up to a normalization factor.
For instance, $ b_H(N)=1, ~b_H({N(N\pm 1) / 2})=2, ~b_H(N^2-1)=0,$ etc. where we have used the dimension to denote the representation. 
(Strictly speaking, the existence of the $\U(1)_{B_H}$ requires any matter with a hidden baryon number $b_H\neq 0$.) 
{We emphasize the hidden baryon number is obtained for the representations solely by the group structure.}

The $\U(1)_{B_H}$ is explicitly broken by a gauge singlet operator $O_{n\neq 0}.$ However, with $n \neq 0$ and $|n_i-\bar{m}_i| \ll N$, the $\ell$ in Eq.~\eq{op} needs to be large to satisfy Eq.~\eq{vnN}. 
This means that $O_{n\neq 0}$ should include many fields, and thus can only appear in the Lagrangian as a high dimensional operator. 
As a result, one gets an arbitrarily precise $\U(1)_{B_H}$ for sufficiently large $N$ and small rank $n_i+ \bar{m}_i$.{ [For $\ab{n_i-\bar{m}_i}\geq N$, one may also get a precise $\U(1)_{B_H}$ with $b_H(r_i)= n_i-\bar{m}_i~({\rm mod}~ N) \OR -( \bar{m}_i-n_i~({\rm mod}~ N))$.]}

In particular, if a large $N$ gauge theory is asymptotically-free, small $n_i$, $\bar{m}_i$ are needed for a negative beta function $\beta(g_N) \equiv \partial g_N / \partial \log(\mu)$ of the gauge coupling $g_N$.
For instance, the beta function with only rank-$k $ representations satisfies
 $16\pi^2\beta/g_N^3+{11\over 3} N \propto N^{k-1}$, which implies only matters with $k=n_i+ \bar{m}_i \leq 2$ are allowed for sufficiently large $N$. 

The $\U(1)_{B_H}$ can be anomalous to other gauge groups if there are chiral fermions.
This is similar to the $\U(1)_B$ - $\SU(2)_L^2$ anomaly induced by the loops of left-handed quarks, i.e., chiral fermions under $\SU(3)_C \times \SU(2)_L$.
Notice that in the presence of other gauge groups, some types of the $O_{n\neq 0}$ are forbidden, and the precision of the $\U(1)_{B_H}$ gets only better. 
As will be discussed, if there are chiral fermions under $\SU(N) \times \SU(3)_C$, there can be an anomaly of $\U(1)_{B_H}$ - $\SU(3)^2_C$.
The anomaly and the good precision of the $\U(1)_{B_H}$ implies that we can identify it as the $\U(1)_{PQ}$.

In a general $\SU(N)$ gauge theory, we obtain the hidden baryon number assignment given in Eq.~\eq{rule}, which is model independent and can lead to a precise $\U(1)_{B_H}$ due to the ``$N$-ality'' of the $\SU(N)$.
In particular, the $\U(1)_{B_H}$ has an arbitrarily good precision for a large $N$ asymptotically-free gauge theory, and it can be identified as the PQ symmetry. 
In our best knowledge, this has not been clearly pointed out in existing literature although many works used such a structure in various forms originating from their choice of gauge symmetries.

\section{\boldmath Models with $\U(1)_{B_H} = \U(1)_{PQ}$}
Here, we address how the $U(1)_{B_H}$ from the $\SU(N)$ can be broken producing a pNGB.
We also show that this pNGB can be identified as a QCD axion.

In order to achieve the $\U(1)_{B_H}$ breaking, there can be many scenarios including
(1) strong dynamics of the $\SU(N)$ gauge theory,
(2) spontaneous breakdown of the $\SU(N)$ with a new Higgs boson, and 
(3) strong dynamics of another gauge group $\SU(M)$.
Our mechanism introduced in the last section is a generic one and does not depend on the breaking mechanism.
In the following, we discuss two scenarios for the sake of concrete model buildings. We will not introduce any scalar field since the mass scale would lead to a fine tuning problem.

Scenario (i): A chiral symmetry breaking of the $\U(1)_{B_H}$ can happen due to the strong dynamics of the $\SU(N)$ gauge theory. 
Let us suppose that $\U(1)_{B_H}$ - $\SU(N)^2$ anomaly is zero.
Otherwise, the pNGB of $\U(1)_{B_H}$ may not appear in the low energy effective theory like the pNGB of the axial $\U(1)$ in the QCD~\cite{Weinberg:1975ui}.
By solving the vanishing conditions for anomalies of $\U(1)_{B_H}$ - $\SU(N)^2$ with Eq.~\eq{rule} and $\SU(N)^3$, up to $n_i+\bar{m}_i\leq 2$, one obtains the chiral fermions as $F_{\rm chi}$ sets of representations of
\beq
\Big({N(N- 1) \over 2}\Big) + 8 N + \Big(\overline{{{N}({N}+ 1) \over 2}}\Big)
\eeq
plus vector-like fermions (and arbitrary scalars in general which but we do not consider).
The 1-loop $\beta$-function of the model is
\beq
\beta(g_N) =- {g_N^3\over 16\pi^2} \Big[\Big({11\over 3}- F_{\rm chi}-\d^{(2)}_{\rm vec}\Big)N -4F_{\rm chi}-\d^{(1)}_{\rm vec}\Big] ,
\eeq
where $\d^{(1,2)}_{\rm vec}$ denote the contributions of the vector-like fermions up to second rank representations, and are positive numbers depending on the contents.
The model can be asymptotically-free if
\beq
N \geq  {12 F_{\rm chi}+3 \d^{(1)}_{\rm vec} \over 11-3F_{\rm chi}-3\d_{\rm vec}^{(2)}}.
\eeq

With $F_{\rm chi}=1$ without a vector-like fermion, the model was studied in Ref.~\cite{Bolognesi:2017pek} in a different context. 
It was discussed that there is a chiral $\SU(8)_F$ flavor symmetry, and  $\SU(8)_F$ can remain unbroken, while the $\U(1)_{B_H}$ is broken due to the strong dynamics of the $\SU(N)$.  

We point out that this $\U(1)_{B_H}$ is at a good precision as shown in the previous section.  Therefore there is a light pNGB. 
If a subgroup of $\SU(8)_F$ is gauged as $\SU(3)_C$ or ${\cal G}_{SM}$, the pNGB may become the QCD axion, since the anomaly of $\U(1)_{B_H}$ - $\SU(8)_F^2$ is non-vanishing.
However several theoretical details are beyond our scope, e.g. whether the $\SU(3)_C$ coupling changes the $\SU(N)$ strong dynamics, and we do not discuss this possibility more.  
{We emphasize that we have shown the group theoretical definition of the $\U(1)_{B_H}$ as Eq.~\eq{rule} makes it easy to identify the model which has an accidental $\U(1)_{PQ}$ at a good precision.}

Scenario (ii):
In what follows, we will consider a more conservative scenario by breaking the $\SU(N)$.  
Even without a scalar field, the $\U(1)_{B_H}$ can be spontaneously broken down by the strong dynamics of another gauge group $\SU(M)$. 
To be conservative, we assume there is no particle both charged under the $\SU(M)$ and ${\cal G}_{SM}$ so that the strong coupling of the SM is irrelevant to the dynamics of the $\SU(M)$.
Moreover, our model would not have a Landau pole of the $\SU(3)_C$ coupling below the Planck-scale~\cite{Dobrescu:1996jp}. 
Because of this assumption we can separate the model and discussion into three sectors: the $\U(1)_{B_H}$ breaking sector where the $\SU(M)$ charged particle lives, the hidden sector where the $\U(1)_{B_H}$ - $\SU(3)_C^2$ anomaly is obtained, and the ordinary SM sector.

\paragraph{$\U(1)_{B_H}$ breaking sector:}
\lac{hlb}
In the $\U(1)_{B_H}$ breaking sector, $\SU(N)$ and $\U(1)_{B_H}$ breaking can take place.
{To see this, consider fermions 
\begin{align}
q: (N, M, 1) , \quad \bar{q}: (\bar{N}, M, 1) , \quad \p_I: (1, \bar{M}, 1) ,
\end{align}
where we denote the corresponding representation of $(\SU(N), ~\SU(M),~{\cal G}_{SM})$, $1$ represents a gauge singlet, and the flavor index $I=1 \cdots 2N$.}
By taking all couplings except for the gauge coupling of $\SU(M)$ to zero, there appears a chiral $\SU(2N)_L \times \SU(2N)_R$ global symmetry.
The chiral symmetry was shown to be broken down when~\cite{Appelquist:1996dq,Appelquist:2009ka}
\beq 
\laq{co4} M \gtrsim N,
\eeq
and one gets the expectation values
\beq
\laq{vev}
\vev{q_\a \p_{I}} \simeq \L_M^3 \delta_{\a I} , \quad \vev{\bar{q}_{\bar{\a}}\p_{I}} \simeq \L_M^3 \delta_{\bar{\a} ,I-N} .
\eeq
Here we have explicitly written down the indices for $\SU(N)$ as $\a$, $\bar{\a}$ and the flavors with a certain field redefinition.
We expect this result does not change when the coupling $g_N$ is small enough, which is what we assume.
Obviously, the $\SU(N)$ and $\U(1)_{B_H}$ are both broken.
The NG modes are $(2N)^2-1$. 
$N^2-1$ NG modes are eaten by the $\SU(N)$ gauge bosons whose mass $\sim g_N \L_M$.
$3N^2-1$ NG modes get masses of order ${g_N\over 4\pi} \L_M$  due to the radiative correction from the $\SU(N)$ gauge interaction, and $1$ is the pNGB of the $\U(1)_{B_H}$.
The decay constants of the pNGBs are
\beq
\laq{fhl}
f={\L_{M} \over  c^{1/3}} ,
\eeq
where $c$ is an $\O(10)$ positive number which increases when $N$ increases for a given $M$~\cite{Appelquist:2009ka}. 
The pNGB of the $U(1)_{B_H}$ has a Planck-scale suppressed mass contribution of $\O(\L_M^d/M_{\rm pl}^{d-2}),$ where $d$ satisfies the condition \eq{fd}.

\paragraph{Hidden sector:}
Let us couple the pNGB of the $\U(1)_{B_H}$ to the SM gauge bosons through the chiral anomaly.
In particular, when there is the $\U(1)_{B_H}$ - $\SU(3)_C^2$ anomaly, the pNGB can be identified as the QCD axion. To this end, let us introduce additional $F_{\rm vis}$ copies of fermions that are chiral under $\SU(N)\times \SU(3)_C,$
\begin{align}
\laq{compo}
\bar{\p}^{\rm vis}: (N, 1 ,\bar{ r}_{\rm SM})  , \quad {{\p}^{\rm vis}_i}:(1,1, {r}_{\rm SM}) , \quad \chi_j: &(\bar{N}, 1,1 ),
 \end{align}
 where the flavor indices $i=1\cdots N \AND j=1 \cdots \dim{[r_{\rm SM}]}.$
Notice that due to the anomaly cancellation of the $\U(1)_Y$ - $\SU(N)^2$,  $r_{\rm SM}$ cannot be a single colored representation with a non-vanishing hypercharge.
By assuming a grand unification theory (GUT) of $ \SU(5)\supset {\cal G}_{SM}$ at around the scale $10^{15-17}\GEV$, it is natural that $r_{\rm SM}$ is in a complete GUT multiplet and thus the anomaly vanishes.
For instance, let us take $r_{\rm SM}=(-2/3_Y, 1_L, \bar{3}_C )+(1/6_Y, 2_L, 3_C )+(1_Y, 1_L, 1_C )$, which form a 10 multiplet of the $\SU(5)$. 
We find that despite the cancellation of the anomaly of $\SU(N)$ - ${\cal G}_{SM}^2$, the anomaly of $\U(1)_{B_H}$ -$\SU(3)_C^2$ is generated.  

The pNGB of the $\U(1)_{B_H}$ from the $\SU(N)$ can be identified as a composite QCD axion~\cite{Kim:1984pt, Choi:1985cb} with a sufficient precision, as we show in the following, if
\beq
\laq{co3}
N\geq 5.
\eeq
Since the strong CP phase is constrained to be $\theta_{CP} \lesssim 10^{-10}$ from the neutron electric dipole moment, the $\U(1)_{PQ}$ should be precise up to the Planck-scale suppressed operator of dimension $\laq{ax} d \gtrsim 9$~\cite{Barr:1992qq}, for the axion decay constant $f_a \gtrsim 10^8\GEV$ from the constraint of Supernova 1987A~\cite{Chang:2018rso}.
(See also~Refs.~\cite{Mayle:1987as, Raffelt:1987yt}.)
For $N=5,$ the lowest dimension breaking term of the $\U(1)_{B_H}$ at the perturbative level is $\chi^5 L_{\rm SM} H_{\rm SM}$, where $L_{\rm SM}$ and $H_{\rm SM}$ are the left-handed lepton fields and Higgs field in the SM, 
respectively.
(When the SM fermions carry some additional gauge charges, this term is forbidden and $N \geq 3$ is allowed while keeping the sufficient quality of the PQ symmetry.)
$\chi^5$ is due to the ``$N$-ality'' while the SM fields are needed for Lorentz invariance. 

The gauge boson coupling of the pNGB in our model is induced by integrating out the $\bar{\p}^{\rm vis}$ and ${{\p}^{\rm vis}_i}$ fermions,  
\beq
\laq{ancp}
\sqrt{N\over 2}{3 F_{\rm vis} \over 16\pi^2}{a\over f}\Big( {5\over 6} g_Y^2 F_Y\tl{F}_Y+
   g_2^2 \tr[F_L\tl{F}_L]+
   g_3^2 \tr[F_C\tl{F}_C]\Big) . 
\eeq
This is the typical coupling of the GUT axion models~\cite{Zhitnitsky:1980tq, Dine:1981rt}. 
Nonetheless, $f$ can be much smaller than the GUT scale without splitting the components of $r_{\rm SM}$ due to a GUT breaking, thanks to the gauge anomaly cancellation of $\U(1)_Y$ - $\SU(N)^2$.
The possibility of the GUT axion-gauge couplings with a smaller decay constant is one of the features of this model.

Note that the fermions of $\bar{\p}^{\rm vis}$, $\and \p^{\rm vis}$ out of the dynamical breaking sector are the main difference from the conventional composite axion models solving the quality problem~e.g. Refs~\cite{ Randall:1992ut, Dobrescu:1996jp, Lillard:2018fdt} (See also Ref.~\cite{Redi:2016esr}).
As we will see, due to this setup the Landau pole problem discussed in Ref.~\cite{Dobrescu:1996jp} can be solved and light exotic fermions are predicted.

Our model can be perturbative up to the GUT scale. 
The 1-loop beta functions of the $\SU(N)$ and $\SU(3)_C$ above the scale $\L_{M}$ are
\begin{align}
\beta(g_N)&= -{g_N^3\over 16\pi^2} \Big({11 \over 3 }N - 2M-10 F_{\rm vis}\Big),\\ 
\beta(g_C)&= -{g_C^3\over 16\pi^2} \(7- 3 F_{\rm vis} N\),
\end{align}
respectively. 
To be consistent with perturbative GUT, the $g_C$ as well as the other SM gauge couplings should be 
perturbative up to the GUT scale, $\sim 10^{16}\GEV.$ This requires 
\beq
\laq{co1}
N F_{\rm vis}\lesssim 10 
\eeq
where we have taken that $g_{C}=0.6$ at $\mu=10^{12}\GEV.$ 
This condition also holds for the $\U(1)_Y \AND \SU(2)_L$ gauge couplings since $\bar{\p}^{\rm vis}\AND \p_i^{\rm vis} $ are in GUT multiplets. 
We found that for $ F_{\rm vis}=1 \OR 2$, the quality of the PQ symmetry can be good enough without violating the perturbativity. 
Although it is not necessary, the $g_N$ can be asymptotically-free for 
\beq
\laq{co2}
N\geq {6\over  11} M+{10\over 11} F_{\rm vis}.
\eeq 
We can easily see the conditions \eq{co1} and \eq{co2} as well as \eq{co4} and \eq{co3} can be simultaneously satisfied.
(The asymptotic freedom of the SM gauge groups may be obtained after they unify into a large gauge group at the GUT scale.)

Because of the chiral property, our model predicts much lighter exotic fermions than the dynamical scale $\L_M$.
The fermions, $ \bar{\p}^{\rm vis} \AND \p_i^{\rm vis}$, get masses through dimension six operators
\beq
\laq{masst}
\bar{\p}^{\rm vis} \p_i^{\rm vis} \tl{y}_{i I} \bar{q}\p_{I} ,
\eeq
where $\tl{y}_{i I}=\tl{y}_i \delta_{i I} $ is dimension $-2$ matrix of flavor, which is diagonalized by the field redefinition. 
{From \Eq{vev}}, this implies that the mass of the fermion is
\begin{align}
\laq{mQ}
&m_{\p^{\rm vis}_i } \simeq \tl{y}_i {\L_M^3} \non \\ 
&\simeq 0.1 {N}^{3\o 2} F^3_{\rm vis} \Big({c\over 60}\Big) \Big(\frac{f_a}{10^{12}\GEV}\Big)^3 \Big(\tl{y}_i^{1\o 2}M_{\rm pl}\Big)^2 \TEV ,
\end{align}
where $f_a \equiv \sqrt{2/N}f/(3F_{\rm vis})$.
We have used \Eq{fhl} and assumed \Eq{masst} is generated through the Planck-scale physics. 
We take the conservative lower bound of $c$~\cite{Appelquist:2009ka}.
The decay constant is constrained from the cosmological axion abundance~\cite{Abbott:1982af,Dine:1982ah,
Preskill:1982cy}, $f_a \lesssim 10^{12}\GEV$, for the Hubble parameter during inflation much greater than the QCD scale~\cite{Graham:2018jyp, Guth:2018hsa}.
In this case, light exotic quarks and leptons, $\psi_{\rm vis} \AND \bar{\psi}_{\rm vis}$,  such as of TeV scale are predicted. 

The lifetimes of these fermions depend on the couplings to the SM particles through either renormalizable or Planck-scale suppressed terms, e.g.
\beq
\tl{c}^i_\tau {H_{\rm SM}} L_{ \tau } \p_{i,e}^{\rm vis},  \quad \tl{{c}}_{t_R} \vev{\bar{q} \p} \bar{\psi}^{\rm vis}_u t_R/ M^2_{\rm pl} ,
\eeq 
where $\tl{c}_\tau \AND \tl{c}_{t_R}$ are dimensionless couplings expected to be $\O(1)$; $L_\t$, $t_R$ are the left-handed $\tau$ lepton and right-handed top quark, respectively; $\p_{i,e}^{\rm vis}$ and $\bar{\p}_u^{\rm vis}$ are the exotic lepton $(1_Y, 1_L, 1_C)$ of $\p_i^{\rm vis}$ and the exotic quark $(2/3_Y, 1_L, 3_C)$ of $\bar{\p}^{\rm vis}$, respectively; similar terms can be written for other components of $\p_i^{\rm vis} \AND \bar{\p}^{\rm vis}$. 
The first term shows that the $\p^{\rm vis}_i$ (and thus $\bar{\p}^{\rm vis}$) can decay into the Higgs and SM fermions if kinematically allowed. 
The second term shows that the mixing term between $\bar{\p}^{\rm vis}$ and the SM fermions is not suppressed compared to the mass term \eq{masst}. 
These facts imply the lifetime of the fermions are much shorter than the cosmological scale, unless we fine-tune the couplings of $\tl{c}^i_\t, \tl{c}_{\tau_R},$ etc. 
On the other hand, depending on the lifetime these visible fermions could be tested at the LHC and future colliders as the fourth (or even higher) generation fermions or vector-like fermions~(see e.g. Refs.~\cite{Ellis:2014dza,Ishiwata:2015cga,Aboubrahim:2018hll,Cacciapaglia:2018lld,Chala:2018qdf}).

Notice that light exotic quarks are predicted independent of the representations of fermions in both the $\U(1)_{B_H}$ breaking sector and hidden sector, when no scalar field charged under the $\SU(N)$ is present. 
Since some quarks are chiral in $\SU(N) \times \SU(3)_C$ to get the pNGB-gluon coupling, 
these quark mass terms are forbidden.  The quark mass is generated  through the breaking of $\SU(N)$ via a term with a dimension at least 6. As a result, the exotic quark masses are around \Eq{mQ}, or smaller.

Finally, let us comment on phenomenological constraints on the model. 
{$\chi_j$ get masses through the dimension 9 term, $(\chi \p q)^2/M_{\rm pl}^5$, and are extremely light.}
Even if very light, ${\chi}_j$ can be cosmologically safe when the reheating temperature of the Universe is sufficiently small and they are not thermalized. 
The thermalization rate $ \sim T^5/f^4$, which is obtained through the heavy $\SU(N)$ gauge boson exchange, implies that the production is ineffective for $T\lesssim  (f^4/  M_{\rm pl })^{1/3}\sim 10^{11}\GEV  \(f/(3\times 10^{12} \GEV)\)^{4/3}$.
The reheating temperature above $\O(10^8) \GEV$ allows thermal leptogenesis~\cite{Fukugita:1986hr} by 
introducing right-handed neutrinos, or leptogenesis through the active neutrino flavor oscillation~\cite{Hamada:2018epb} with certain inflaton decay channels by assuming any mechanisms generating the $(L_{\rm SM} H_{\rm SM})^2$ term. (See e.g. Ref.~\cite{Yin:2018qcs}.) 
Such a low reheating temperature is consistent with the low-scale inflation which avoids the generation of domain-wall and isocurvature perturbation constraint on the axion dark matter.

\section{Summary and Discussions}
In this work, we have discussed a natural and general possibility to provide the QCD axion that can solve the strong CP problem. 
We have focused on dynamical breaking of the {$\U(1)_{B_H}$} for the concrete examples to realize our idea.
Yet, one can also introduce scalar fields of {certain} representations under the $\SU(N)$ for the symmetry breaking. 
In both cases, some hidden gauge fields can be light or even massless depending on the breaking scenario. These gauge bosons may have a dark axion portal interaction, such as an axion-photon-dark photon vertex, resulting in various new phenomena~\cite{Kaneta:2016wvf, Choi:2016kke, Daido:2018dmu}.

A product group of $\SU(N)\times \U(1)'$, where $\U(1)'$ is a gauge symmetry or a good global symmetry, is also an interesting setup, where arbitrary combinations of $\U(1)_{ B_H} \AND \U(1)'$ become good symmetries (even with better quality than $\U(1)_{B_H}$).
For instance, particles charged under $\U(1)'$ gauge symmetry but not charged under $\SU(N)$ may carry an accidental global symmetry originated from $\U(1)_{B_H}.$
 
From the analogy of the baryon number in the SM, we have shown that a hidden baryon number of the $\SU(N)$ gauge theory can be a natural candidate for the charge of the PQ symmetry. 
The hidden baryon number assignment is solely determined by the structure of the $\SU(N)$ gauge group.
The PQ symmetry can be {arbitrarily precise} when the $\SU(N)$ gauge theory is asymptotically-free  with large enough $N$.

As an interesting phenomenology, in our composite axion model based on the $\U(1)_{B_H}$, exotic quarks relevant to the axion-gluon-gluon coupling can remain light enough to be tested in the present and future experiments.
Many other interesting phenomenology or model buildings based on our mechanism are warranted.
For instance, the accidental $\U(1)_{B_H}$ can be also used for providing an ALP, or purely a precise global continuous symmetry with the charge assignment \eq{rule}. 
In particular, a precise $\U(1)_{B_H}$ is attractive for the ALP as the unified inflaton and dark matter candidate~\cite{Daido:2017wwb, Daido:2017tbr}.

\acknowledgments{
This work was supported by National Research Foundation (NRF) Strategic Research Program (NRF-2017R1E1A1A01072736).
}

\end{document}